\documentclass[%
 reprint,
 amsmath,amssymb,
 aps,
 prx,
]{revtex4-2}
\usepackage{graphicx}
\usepackage{dcolumn}
\usepackage{bm}
\usepackage{braket}
\usepackage{booktabs}
\usepackage{xcolor}
\usepackage{siunitx,xfrac}
\usepackage[normalem]{ulem}  
\DeclareSIUnit{\sqrthz}{\sqrt{\hertz}}
\DeclareSIUnit{\dBm}{dBm}
\DeclareSIUnit{\dBc}{dBc}
\DeclareSIUnit{\dBi}{dBi}
\DeclareSIUnit{\Vcm}{(V/cm)}
\DeclareSIUnit{\cmV}{(cm/V)}
\DeclareSIUnit{\uVcm}{(\micro V/cm)}
\DeclareSIUnit{\uVm}{(\micro V/m)}
\DeclareSIUnit{\Vm}{(V/m)}
\DeclareSIUnit{\mVm}{(mV/m)}

\begin{document}

\preprint{APS/123-QED}

\title{Waveguide-coupled Rydberg spectrum analyzer from 0 to 20~GHz}

\author{David H. Meyer}
\author{Paul D. Kunz}%


\author{Kevin C. Cox}
\email[Corresponding author: ]{kevin.c.cox29.civ@mail.mil}

\affiliation{
 CCDC US Army Research Laboratory, Adelphi, MD 20783 USA
}%

\date{\today}

\begin{abstract}
We demonstrate an atomic radio-frequency (RF) receiver and spectrum analyzer based on thermal Rydberg atoms coupled to a planar microwave waveguide.  We use an off-resonant RF heterodyne technique to achieve continuous operation for carrier frequencies ranging from DC to \SI{20}{\giga\hertz}.  The system achieves an intrinsic sensitivity of up to \SI{-120(2)}{\dBm\per\hertz}, DC coupling, \SI{4}{\mega\hertz} instantaneous bandwidth, and over \SI{80}{\deci\bel} of linear dynamic range.  By connecting through a low-noise preamplifier, we demonstrate high-performance spectrum analysis with peak sensitivity of better than \SI{-145}{\dBm\per\hertz}.  Attaching a standard rabbit-ears antenna, the spectrum analyzer detects weak ambient signals including FM radio, AM radio, Wi-Fi, and Bluetooth.  We also demonstrate waveguide-readout of the thermal Rydberg ensemble by non-destructively probing waveguide-atom interactions.  The system opens the door for small, room-temperature, ensemble-based Rydberg sensors that surpass the sensitivity, bandwidth, and precision limitations of standard RF sensors, receivers, and analyzers.
\end{abstract}

\maketitle

\section{Introduction\label{sec:Intro}}

Sensors based on quantum constituents have unique properties that distinguish them from traditional technologies.  The absolute sameness of quantum particles often leads to exquisite precision, and their response and performance are accurately linked to first-principle predictions.  Quantum sensors of time (atomic clocks) and magnetic fields (magnetometers) have achieved record performance, and other classes of quantum sensors are expected to follow.

Quantum sensors for radio-frequency (RF) electro-magnetic fields are a swiftly emerging subset, and will be critical in the future, as ever-increasing networking and informational demands require greater capabilities to utilize the finite spectrum.   But non-cryogenic quantum RF sensors do not currently match the sensitivity of traditional receivers that use standard electronics \cite{meyer_assessment_2020}.  And, until now, individual quantum RF sensor measurements have only covered small portions of the spectrum.

Here, we present a near-room-temperature RF quantum sensor based on thermal Rydberg atoms that operates continuously from 0 to \SI{20}{\giga\hertz}, and rivals the performance of commercially-available spectrum analyzers with high sensitivity, \SI{4}{\mega\hertz} instantaneous bandwidth, and over \SI{80}{\deci\bel} of linear dynamic range.  Our sensor improves upon previously demonstrated Rydberg RF sensors, by 1) improving sensitivity to RF power by confining the RF field in a small mode volume that closely matches the sensing volume and 2) utilizing an off-resonance heterodyne technique to greatly boost the sensitivity at arbitrary frequencies, far from a resonant Rydberg transition.  The system can be operated with or without a low-noise preamplifier for the input RF signals, with numerous possible paths for miniaturization and improved performance.  

\section{Prospects for Rydberg sensors}
Overall, there are several reasons for excitement about Rydberg RF sensors. Being a quantum sensor that measures quantum phase accumulation of an atomic state, they are expected to surpass several foundational limitations to traditional receivers. 
First, the Rydberg state's response is linked to fundamental constants and is easily calculated to high accuracy \cite{sedlacek_microwave_2012, holloway_broadband_2014}, meaning the receiver can serve as an absolute calibration over a large, technologically-relevant parameter space, for frequencies from DC up to \SI{1}{\tera\hertz} \cite{gurtler_imaging_2003,downes_full-field_2020, meyer_assessment_2020}. 
Second, active quantum sensors may achieve high bandwidths independent of the carrier frequency and are not, in general, subject to the bandwidth limitations of passive receivers using resonant electrically-small antennas \cite{cox_quantum-limited_2018}.  
Third, and arguably most exciting, is the possibility to avoid internal thermal (Johnson) noise, even at room temperature.  This is possible since the Rydberg sensor relies on measuring the atoms' quantized internal states with low-entropy laser beams.

These aspects and overall performance of Rydberg RF sensors have been explored in increasing depth in recent years.  Several key experiments identified the usefulness of Rydberg atoms for electric field sensing \cite{osterwalder_using_1999, gurtler_imaging_2003, mohapatra_giant_2008, sedlacek_microwave_2012, holloway_broadband_2014}.   Recent demonstrations have observed high precision for electric field calibration \cite{sedlacek_microwave_2012, holloway_broadband_2014}, terahertz imaging \cite{gurtler_imaging_2003,downes_full-field_2020}, sensing very strong fields \cite{paradis_atomic_2019},  proof-of-principle communication reception \cite{meyer_digital_2018, deb_radio-over-fiber_2018, anderson_atomic_2018, jiao_atom-based_2019}, operation at low frequencies \cite{jau_vapor-cell-based_2020,  cox_quantum-limited_2018}, and entanglement-enhanced E-field sensing \cite{facon_sensitive_2016}.  Perhaps the most seminal achievement of the decade used a beam of Rydberg atoms traversing a superconducting microwave cavity to create and stabilize non-classical states of light \cite{sayrin_real-time_2011, raimond_manipulating_2001}.  

\begin{figure*}[t]
\centering
\includegraphics[width=\textwidth]{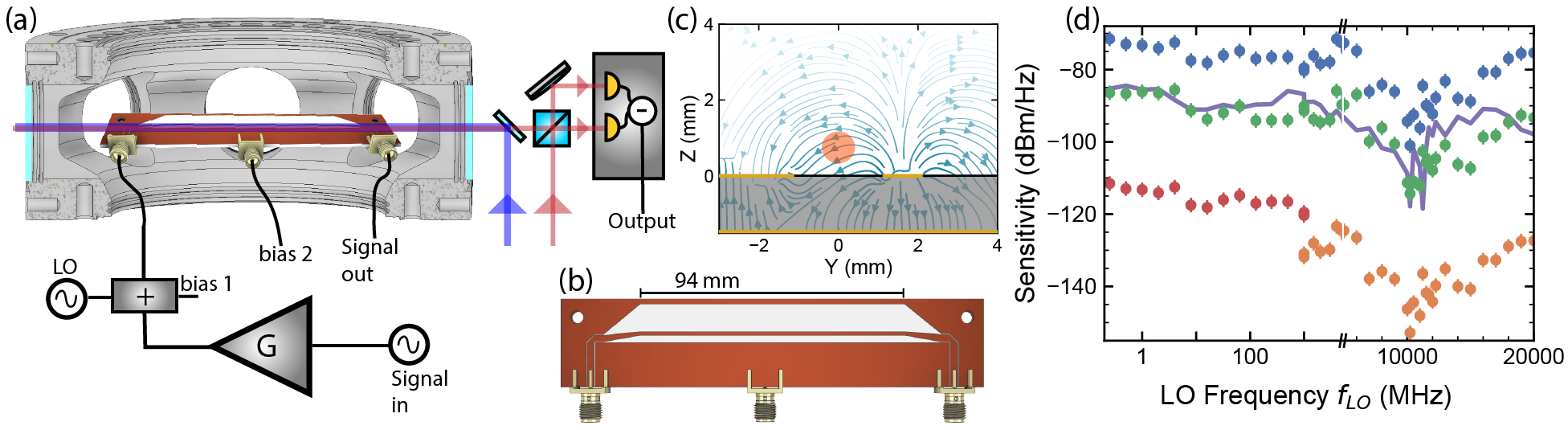}
\caption{ (a)  Simplified experimental setup.  Rydberg atoms are detected using a \SI{780}{\nano\meter} and \SI{480}{\nano\meter} beam, counter-propagating above a microwave circuit, with connections for input, output, and DC bias. Signals are detected in homodyne, using a balanced detector.  (b)  Microwave circuit.  A coplanar waveguide transitions to a circuit region where the evanescent electric field area is matched to the Rydberg interrogation area.  A third SMA cable, in the center, is used to add a DC bias to the circuit backplane, to help zero ambient DC fields.  (c) Field simulations of the microwave circuit along a middle slice of the board. Atoms are interrogated over a \SI{2}{\milli\meter} gap between the signal conductor and ground conductor. (d) Measured Sensitivity versus frequency.  Directly measured (blue), intrinsic PSN-limited (green), and directly measured using preamplification (red and orange) values are displayed. All error bars represent \SI{2}{\deci\bel} total estimated standard deviation due to known statistical and systematic factors. The theoretically modelled intrinsic Sensitivity, with no preamplification, is shown as a purple line. Note that lower/high frequencies are shown on log/linear scales, respectively.}
\label{fig:main}
\end{figure*}

Despite this rapidly increasing, and exciting, body of work, additional advances in Rydberg sensor performance are required before they become useful as state-of-the-art, RF receivers.  For one, prior works have been confined to frequencies near resonant transitions between Rydberg states. While there are hundreds of these potential transitions, they are unevenly distributed between $\sim1$ and \SI{1000}{\giga\hertz} with highly variable sensitivity \cite{meyer_assessment_2020}. Using all of the available transitions requires a high-power laser system agile enough to rapidly tune over multiple nanometers while maintaining narrow linewidth and high precision frequency stabilization to atomic transitions.  Further, previous room temperature experiments have only been sensitive enough to detect transmitted RF fields that are significantly stronger than most real-world signals.  This is partially because current free-space sensors have inherently weak coupling to incoming RF and microwave modes, and a sensing area that is significantly smaller than the diffraction limit (order $\lambda^2$) in most cases. Even demonstrations with decent electric field sensitivity have detected signals from RF transmitters that were driven with relatively high power at their input and/or from horns a few centimeters away from the sensing volume. Though several other recent experiments have placed thermal Rydberg atoms near RF and microwave structures for sensing purposes, wideband operation and RF power sensitivity were not directly reported \cite{fan_subwavelength_2014,holloway_quantum-based_2018, simons_fiber-coupled_2018, simons_embedding_2019}. Taken together, these realities have made it difficult to realize a continuous, wideband Rydberg sensor of weak RF fields.

To alleviate the poor coupling of free space devices, our system receives electric fields into a planar microwave waveguide that concentrates the electric field into a sub-wavelength region.  Rydberg atoms are created and probed directly over the gap between a signal trace and ground plane, where the evanescent electric field is concentrated to a few square millimeters. This waveguide-coupled Rydberg sensor is not likely to achieve absolute electric field accuracy on par with previous free-space Rydberg electric field measurements. However, it greatly improves sensor sensitivity, a critical metric for large classes of RF applications in communications, sensing, and spectrum awareness.

We also utilize an off-resonant heterodyne technique recently reported in Reference \onlinecite{jau_vapor-cell-based_2020}, combining the input RF signal with a strong RF local oscillator.  This enables wideband operation with no tuning of the Rydberg laser by taking advantage of the square-law response of the off-resonant atomic light shift. Heterodyning increases the response at an arbitrary selected RF input frequency and linearizes the sensor (see App. \ref{app:RydResponse} for details).  Several recent experiments also utilized local oscillators on resonance \cite{gordon_weak_2019, jing_atomic_2020} to sense free-space electric fields, where the Rydberg response is linear to the applied field.

Altogether, we achieve continuous operation with good sensitivity continuously from DC to \SI{20}{\giga\hertz}, where the high-frequency limit of our system is primarily due to the limited bandwidth of the vacuum feedthroughs and microwave circuit, both of which can be improved in future rounds of engineering.

\section{Sensor performance}

A diagram of the apparatus is shown in Figure \ref{fig:main}.  A blue \SI{480}{\nano\meter} ``Rydberg coupling'' laser and a near-infrared \SI{780}{\nano\meter} ``probe'' laser beam counter-propagate through a rubidium-filled vacuum chamber (Fig.~\ref{fig:main}(a)).  A wideband microwave waveguide is mounted inside the chamber, shown in Fig.~\ref{fig:main}(a) and (b).  The beams excite Rydberg atoms directly over the waveguide, to the $\ket{n=59,D_{5/2}}$ state, in a region where the evanescent RF mode is approximately matched to the size of the optical beams (Fig.~\ref{fig:main}(c)). While the evanescent electric field distribution of our coplanar waveguide does not have a simple algebraic form, the mode area above the gap between the signal and ground traces is of order the gap width squared. Signal input power-to-evanescent field conversion factors can be determined empirically or calculated numerically (as described in App.~\ref{app:WaveguidePerf}). Signals on the RF waveguide perturb the energy of the Rydberg state, and the shifts are observed using electro-magnetically induced transparency (EIT) spectroscopy.  As such, RF signals are transduced into optical signals that are measured via optical homodyne on a balanced photo-detector.  On the RF waveguide input, a local oscillator (subsequently referred to as, ``LO'') is combined with the weak RF signal $S_\text{in}$.  The LO improves the sensor sensitivity when the RF fields are off-resonant from a dipole-allowed Rydberg transition, the nearest for our target Rydberg state being at \SI{10.223336}{\giga\hertz}, because the induced time-averaged (in detection timescale $\tau$) light shift is $s \propto \langle E^2 \rangle_\tau \approx E_\text{LO} E_\text{S}$ for electric field $E$ consisting of a local oscillator (LO) and a Signal (S) component.  Additionally, we add two DC bias voltages (bias 1 and bias 2 in Fig.~\ref{fig:main}(a)) to zero the DC field at the atoms' location.  Bias 1 is applied to the center signal conductor, and bias 2 is applied to the backplane of the circuit board. Both voltages are necessary to provide sufficient cancellation of the DC field in the 2D plane perpendicular to the waveguide propagation axis. Initial preamplification of the input signal may be achieved with an amplifier of gain $G$. Additional details about the apparatus are available in App.~\ref{app:ExpDetails}.

\begin{figure}[t]
\centering
\includegraphics[width=\columnwidth]{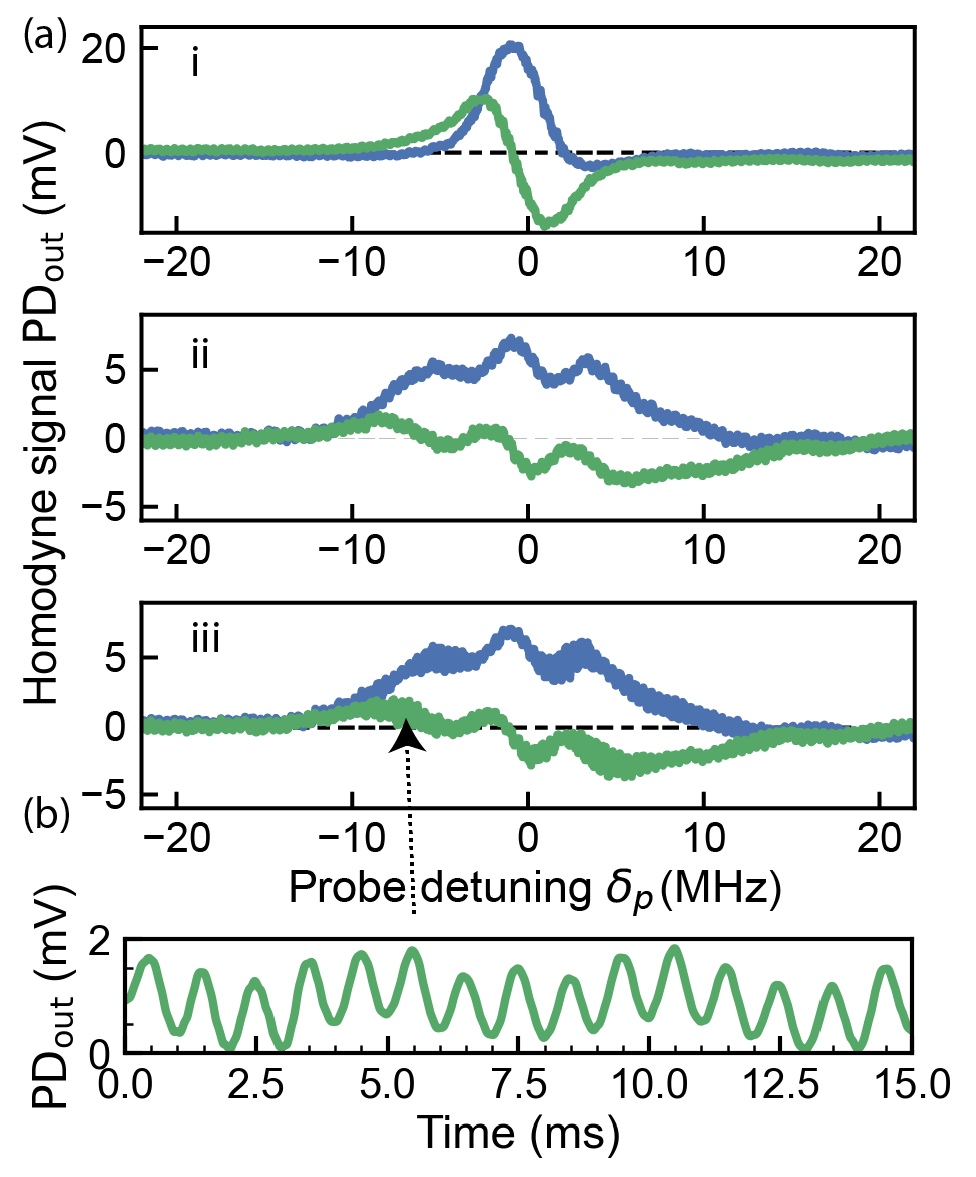}
\caption{Sensor signals. (a) Detected signals versus probe detuning; optical amplitude quadrature in blue, phase quadrature in green, for (i) no microwaves, (ii) applied microwave LO at $f_\text{LO}=\SI{10.2233}{\giga\hertz}$, but no signal field, and (iii) applied LO and signal field with offset $\delta = \SI{1}{\kilo\hertz}$.   (b) Exploded view of (a)iii showing \SI{1}{\kilo\hertz} modulation.}
\label{fig:SensorSigs}
\end{figure}

We plot the resulting sensitivity of the system to RF signals in Fig.~\ref{fig:main}(d).  The directly measured Sensitivity (equivalent here to the minimum detectable signal in a bandwidth of \SI{1}{\hertz}), is shown from \SI{100}{\kilo\hertz} to \SI{20}{\giga\hertz} as the blue points. The circuit is DC-coupled and the sensitivity extends to arbitrarily low frequencies below \SI{1}{\kilo\hertz}, overcoming a common problem with previous vapor cell experiments \cite{cox_quantum-limited_2018, jau_vapor-cell-based_2020}.  The system is fundamentally limited by photon shot noise (PSN) in the optical homodyne readout that is correlated to atomic wave-function collapse \cite{cox_quantum-limited_2018}.  However, over the heterodyne operation bandwidth (DC to approximately \SI{10}{\mega\hertz}), there is additional readout noise due to residual, uncancelled phase noise in the probe laser (see App.~\ref{app:ReadoutSens}).  With additional engineering work, we expect that this noise can be eliminated, leading to an intrinsic photon shot noise (PSN) limited Sensitivity (after independently calibrating and subtracting the additional phase noise in readout) shown as green points in Fig.~\ref{fig:main}(d).  All sensitivities are referred to the input SMA connector of the microwave circuit, by independently measuring losses in the input cables and input vacuum feedthrough. For each frequency, we optimized the input LO power for peak Sensitivity.  We find optimum values, measured at the waveguide input, ranging from \SI{-22.7}{\dBm} at the \SI{10.223}{\giga\hertz} resonance up to \SI{13.7}{\dBm} far from the resonances.  Further increases in LO power at each point lead to broadening of the spectroscopy peak.  We expect that this is due to RF field inhomogeneity, splittings of Rydberg Zeeman sublevels, and/or interactions with other states in the Rydberg manifold.  Future work will be required to understand what further sensitivity gains may be realized using RF heterodyne.

\begin{figure}[t]
\centering
\includegraphics[width=\columnwidth]{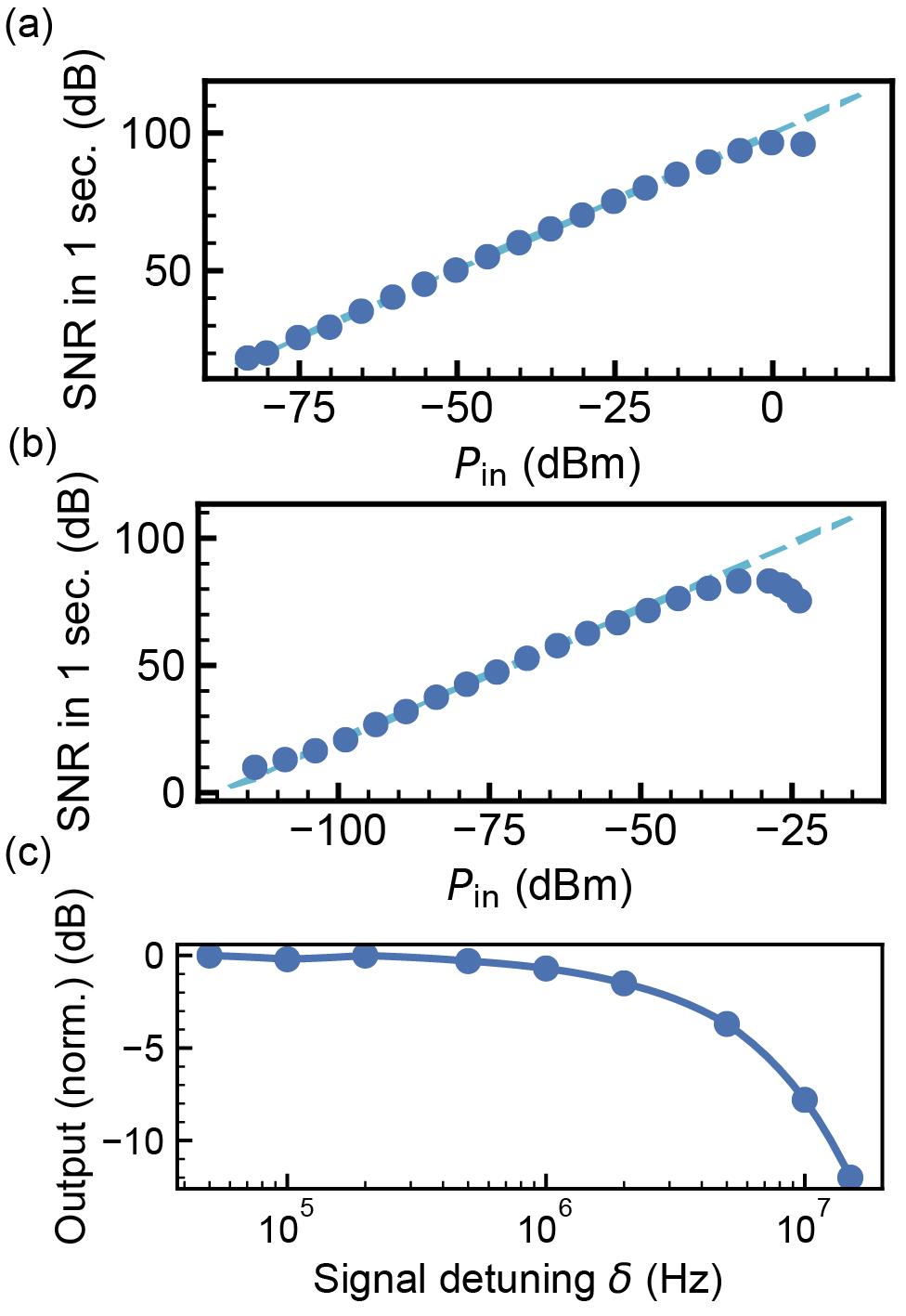}
\caption{Sensor performance (a) Linear dynamic range for $f_\text{LO} = \SI{2}{\giga\hertz}$, $\delta = \SI{700}{\kilo\hertz}$.  In a \SI{1}{\second} measurement, the sensor responds linearly over \SI{80}{\deci\bel}.  PSN-limited SNR is plotted.  (b) Sensor response with $f_\text{LO} = \SI{10.2233}{\giga\hertz}$, $\delta = \SI{700}{\kilo\hertz}$.  (c)  Output signal versus instantaneous frequency $\delta$.  The receiver has a \SI{3}{\deci\bel} reduction at \SI{4.0}{\mega\hertz}.}
\label{fig:SensPerform}
\end{figure}

We also plot Sensitivity using preamplification.  On the low-frequency half of the plot (red), we consider two stages of gain using Minicircuits ZFL-1000LN amplifiers, and on the high frequency half (orange), two Minicircuits ZVA-213-S amplifiers \footnote{References to any specific commercial products do not constitute an endorsement or recommendation by the U.S. Government or the U.S. Army. They are provided solely to serve in the interest of completeness.}.  The resulting Sensitivity is calculated by input referring the intrinsic noise (green) and combining with the input noise of the preamplifiers.  This improves Sensitivity by more than \SI{40}{\deci\bel} beyond the directly observed Sensitivity.  More advanced/optimized preamplification schemes are also possible that would allow the system to reach the thermal noise limit of the amplifiers at approximately \SI{-171}{\dBm\per\hertz}. 

We plot the theoretically modelled intrinsic Sensitivity of the device as a purple line in Fig.~\ref{fig:main}(d).  The theory is calculated using a Floquet analysis that calculates the Rydberg response to an arbitrary RF electric field \cite{meyer_assessment_2020}. The input RF LO and Signal powers are combined with a finite-element calculation using COMSOL Multiphysics modelling software, to determine the electric field at the location of the atoms for an applied voltage (see Appendices \ref{app:WaveguidePerf} and \ref{app:Floquet} for details). We determine Sensitivity from the calculated signal response and the PSN-limited noise.
The largest deviations between the prediction (purple) and measured data (green) are above \SI{10}{\giga\hertz}, where the waveguide performance begins to deteriorate. The two sharp peaks in Sensitivity correspond to the two nearest dipole transitions from the $\ket{59D_{5/2}}$ state at 10.2233 \& \SI{11.2258}{\giga\hertz}.

Typical output signals from the apparatus are shown in Figure \ref{fig:SensorSigs}, as a function of the probe laser's optical detuning from EIT resonance, labelled $\delta_p$.  When no LO or Signal RF fields are applied (a), we observe a bare EIT spectroscopy signal in the optical amplitude quadrature (blue) or phase quadrature (green).  By applying the static LO, the spectroscopic signals are perturbed due to Autler-Townes splitting (near the \SI{10.223336}{\giga\hertz} resonance of the $\ket{59D_{5/2}}$ to $\ket{60P_{3/2}}$ transition, as in Fig.~\ref{fig:SensorSigs}(b)) or AC Stark shifts (when off resonance).  By applying signal $S_\text{in}$, with frequency $f_\text{S}$ near $f_\text{LO}$, we induce fluctuations in the output.  In Figure \ref{fig:SensorSigs}(c), the signal is applied at a detuning of $\delta = \SI{1}{\kilo\hertz}$, where this detuning is defined as $\delta=f_\text{S}-f_\text{LO}$ and $|\delta|$ is equivalent to the offset (or ``baseband'') frequency in the optical Output spectrum.
The applied signal results in the additional fuzz in the trace, with exploded view in Fig.~\ref{fig:SensorSigs}(d) showing the \SI{1}{\kilo\hertz} beat.  Given that the shape of the EIT spectroscopy signal, and therefore the peak sensitivity point, is governed by the strong LO field,
$\delta_p$ should be adjusted for different LO powers to achieve peak sensitivity. 

One key advantage of the Rydberg sensor is that RF power may be measured dispersively, \emph{i.e.}~not absorbed into the atoms.  A technical advantage of this is that the sensor can exhibit extremely high dynamic range and is not adversely affected by high input power or DC offsets (contrary to most sensitive spectrum analyzers).  Recent work has demonstrated using Rydberg atoms to detect local electric fields of over \SI{5}{\kilo\volt\per\meter} \cite{paradis_atomic_2019}.  

We demonstrate the sensor's high dynamic range by measuring the Signal-to-Noise Ratio (SNR) as a function of the input Signal power, $P_\text{in}$.
Figure \ref{fig:SensPerform}(a) plots the PSN-limited SNR of the sensor as a function of signal input power for a \SI{2}{\giga\hertz} signal.  The LO power is constant for this data (\SI{3.9}{\dBm} at the waveguide input), with $\delta = \SI{700}{\kilo\hertz}$ and the signal response is linear over an input power from \SI{-85}{\dBm} to \SI{-5}{\dBm}.  This data was measured in a \SI{1}{\hertz} bandwidth and the results are consistent with sensitivity measurements taken at higher bandwidths up to \SI{1}{\mega\hertz}.  

\begin{figure}[t]
\centering
\includegraphics[width=\columnwidth]{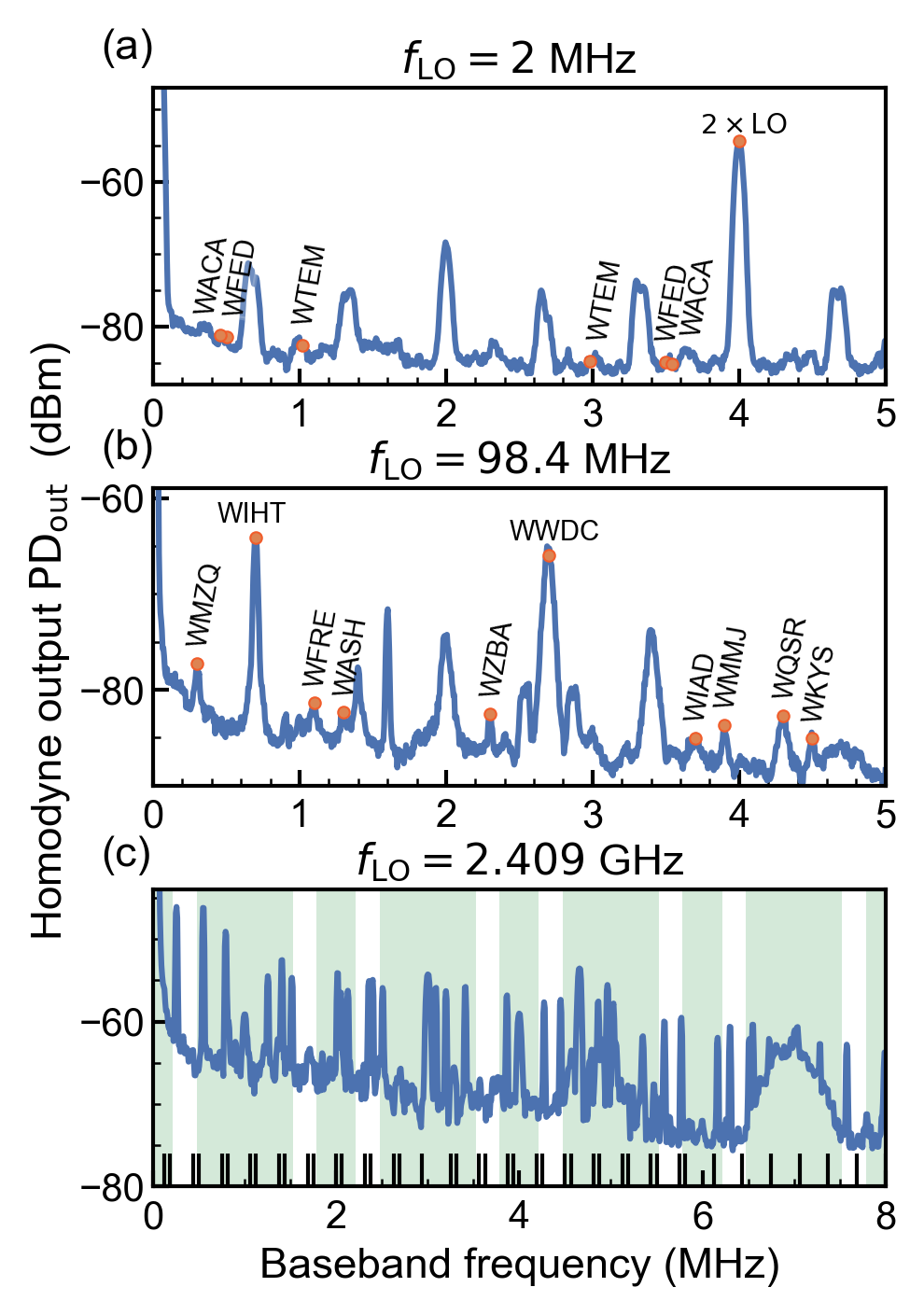}
\caption{RF signals observed inside the lab building using a rabbit-ears antenna.  We tune the LO to $f_\text{LO} = \SI{2}{\mega\hertz}$, \SI{98.4}{\mega\hertz}, and \SI{2.409}{\giga\hertz} to observe AM radio stations, FM radio stations, WLAN, and Bluetooth signals.  (a) AM radio stations sampled around \SI{2}{\mega\hertz}. The doubled LO is directly observed at \SI{4}{\mega\hertz}, since $f_\text{LO}$ is within the instantaneous bandwidth. Other large peaks are spurious signals from lab equipment.  (b) FM stations sampled instantaneously around \SI{98.4}{\mega\hertz}.  (c) WLAN and Bluetooth signals detected over the course of several minutes around \SI{2.409}{\giga\hertz}.  The data was sampled in a max-hold configuration, showing packets sent over many WLAN subchannels (black ticks) and Bluetooth bands (highlighted in green).}
\label{fig:Signals}
\end{figure}

SNR versus input power for a resonant signal of $f_\text{LO} = \SI{10.2233}{\giga\hertz}$, $\delta = \SI{700}{\kilo\hertz}$, and \SI{-36}{\dBm} of LO power at the waveguide input is shown in Fig.~\ref{fig:SensPerform}(b).  A similar dynamic range is observed.  It is again worth reiterating that there is no nearby damage threshold for the apparatus or point where the system physics changes dramatically. By adjusting optical detunings, higher dynamic ranges are feasible.  The data in Fig.~\ref{fig:SensPerform}(b) confirms the  PSN-limited input Sensitivity of \SI{-120(2)}{\dBm\per\hertz} (\SI{-101(2)}{\dBm\per\hertz} directly measured) \footnote{All error bars in plots and parenthetical notation represent total estimated standard deviation due to known statistical and systematic factors}.  
Figure \ref{fig:SensPerform}(c) is a measurement of the sensor response (normalized to 1) as a function of detuning $\delta$.  The \SI{3}{\deci\bel} instantaneous bandwidth is \SI{4.0}{\mega\hertz}, governed by the EIT bandwidth \cite{meyer_digital_2018}.  This bandwidth is independent of carrier frequency, and may be improved in the future with alternative readout schemes. 

\section{Sensing ambient signals}

The increased sensitivity and wide tuning range allows us to detect ambient RF signals in multiple bands inside our lab using a standard rabbit ears antenna connected to the Rydberg receiver.  At each LO frequency, $f_\text{LO}$, we sample data directly from the optical readout and record the resulting spectra. In Fig.~\ref{fig:Signals} we show the spectra from measurements in the AM, FM and \SI{2.4}{\giga\hertz} Industrial, Scientific, and Medical (ISM) bands. For the AM and FM bands, we attached the antenna to the inside of an outward facing window and connected it to the input of the analyzer using two preamplifiers (with combined gain of \SI{60}{\deci\bel}) and a 50 foot BNC cable. The building attenuates the AM/FM signals by \SI{30}{\deci\bel} relative to standing 50 meters away. 
In the ISM band, the antenna was placed within the lab near the experimental apparatus, using a single preamplifier with \SI{38}{\deci\bel} of gain.

In Fig \ref{fig:Signals}(a), the LO frequency is $f_\text{LO} = \SI{2}{\mega\hertz}$, lying within the instantaneous bandwidth of the sensor.  This leads to a large signal at \SI{4}{\mega\hertz}.  Known AM radio stations are labeled with orange dots.  We verified the station AM \SI{980}{\kilo\hertz} (WTEM) with audio playback using an AM demodulator on the Output.  Other large peaks are believed to be spurious signals generated from lab electronics, and their spectral reflections/harmonics.  

Figure \ref{fig:Signals}(b) shows the spectrum around \SI{98.4}{\mega\hertz}, including a number of FM radio stations.  For this data, and higher frequency data, the LO and its harmonics are outside of the instantaneous bandwidth, and are therefore not observed.  The strongest station is at \SI{101.1}{\mega\hertz} (WWDC), having visible digital FM sidebands.  

Figure \ref{fig:Signals}(c) shows a max-hold sampling around \SI{2.409}{\giga\hertz} of Wi-Fi packets broadcast on channels 1-3 of the 802.11 WLAN standard and Bluetooth channels spanning \SIrange{2.402}{2.417}{\giga\hertz}.  These packets were sampled over the course of several minutes, with several active computers and phones in the area.  Each pulse represents either an orthogonal frequency domain multiplexing (OFDM) subcarrier from a Wi-Fi packet, or a frequency shift keying (FSK)/phase shift keying (PSK) pulse from a Bluetooth packet. Baseband frequency locations of the Wi-Fi subcarriers are denoted by black tick marks along the bottom axis and Bluetooth channels are denoted by green bands.  The highly dense band around \SI{7}{\mega\hertz} corresponds to the \SI{2.402}{\giga\hertz} Bluetooth advertising channel.  

\section{Microwave detection of Rydberg atoms}

The atom-circuit coupling is large enough in this experiment to directly detect the presence of the Rydberg atoms by weakly interrogating the microwave waveguide.  This demonstration is an initial step to extending the seminal work of microwave-Rydberg quantum electrodynamics \cite{sayrin_real-time_2011} to the room-temperature regime.  Further, Rydberg-waveguide readout will be a useful tool for future sensor iterations.  For one, direct microwave readout of Rydberg populations is not sensitive to Doppler shifts that plague current optical readout schemes.  Second, strong collective coupling between the Rydberg ensemble and the microwave circuit may lead to a number of impactful experimental possibilities, such as quantum frequency conversion between optical and RF signals, entanglement generation, and/or Rydberg masers.  

\begin{figure}[t]
\centering
\includegraphics[width=\columnwidth]{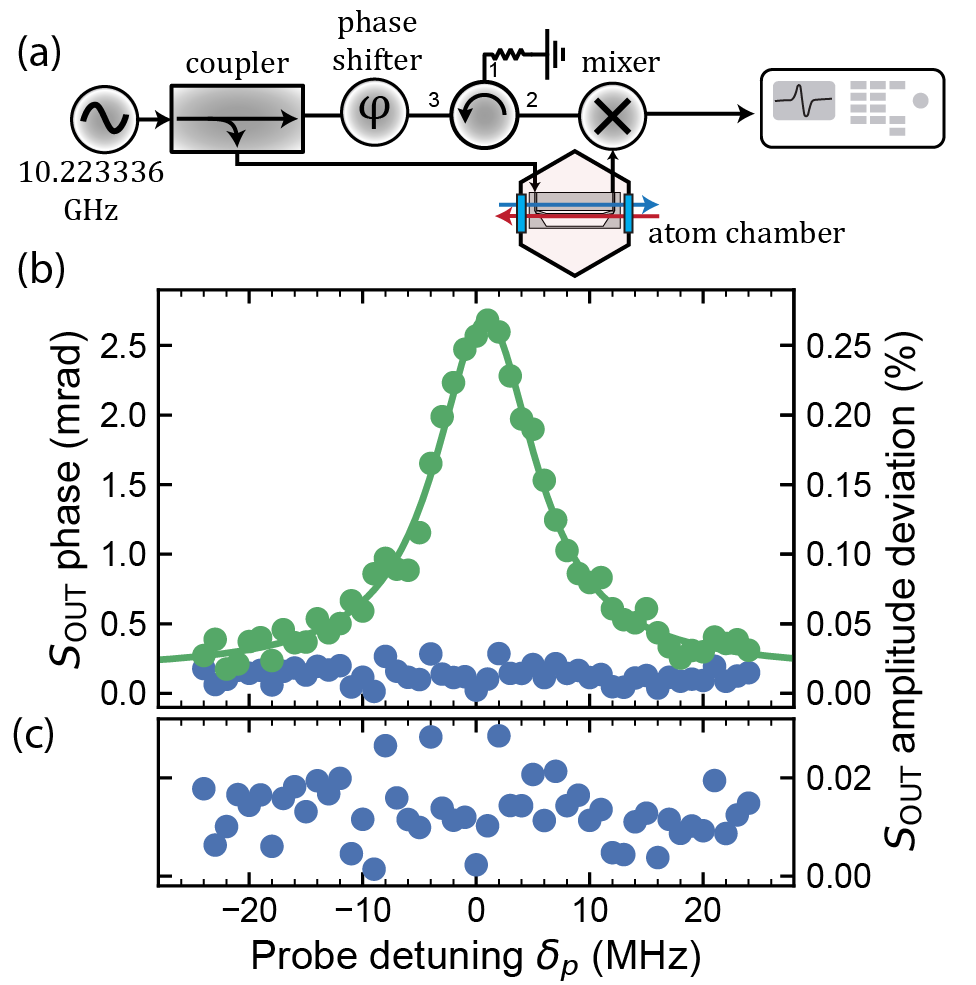}
\caption{Microwave readout of Rydberg atoms. (a) Circuit diagram for microwave homodyne detection at \SI{10.223}{\giga\hertz}.  
(b) Amplitude (blue) and phase (green) measurements of the microwaves transmitted through the chamber, versus probe detuning.  The \SI{3}{\milli\radian} phase deflection, lorentzian about Rydberg resonance, indicates the presence of Rydberg atoms. The lack of signal in the amplitude measurement indicates the Rydberg atoms do not absorb the field.  (c) Zoomed view of the amplitude data in (b).}
\label{fig:MWreadout}
\end{figure}

To observe the Rydberg-circuit coupling, we construct an additional microwave homodyne setup, shown in Figure \ref{fig:MWreadout}(a), to precisely detect the phase of the output microwave signal $S_\text{out}$ (that were simply terminated in the previous measurements).  The atom-induced microwave phase can be readily observed by sweeping the probe laser detuning (Fig.~\ref{fig:MWreadout}(b)) across the dipole-allowed Rydberg transition at \SI{10.2233}{\giga\hertz}.  The atoms present a \SI{3}{\milli\radian} deflection (green points) on resonance, yielding a Lorentzian-shaped signal (green fit).  By tuning the microwave homodyne detection to the amplitude-sensitive quadrature, we confirm that the Rydberg atoms do not significantly absorb the microwaves (blue points of Figs.~\ref{fig:MWreadout}(b-c)). The percent amplitude deviation of the homodyne signal, relative to the total homodyne fringe height in both quadratures, is shown on the right axis.  The measurements of Fig.~\ref{fig:MWreadout} indicate weak coupling.  However, future room-temperature experiments should be expected to reach collective cooperativity greater than 1.

\section{Discussion}

Our results achieve an increased performance level for thermal Rydberg sensors, demonstrating continuous operation over a large frequency range and detecting weak, real-world RF signals.  But the experiment and data also point a clear direction for further improvements.  Circuit-atom coupling may be improved with a lower dielectric substrate, more sophisticated waveguide design, or a tunable resonant circuit. These improvements would increase the electric field strength above the planar surface, increasing the RF-atom coupling. However, one clear advantage of the current non-resonant design is the ease of wideband operation, which would be lost in a resonant design.  

Using appropriate preamplifiers, the present experimental configuration can match the sensitivity of standard spectrum analyzers and receivers, with small positive noise figure.   However, we re-emphasize that with continuing effort, the Rydberg platform may be expected to significantly outperform most wideband RF sensors, with no preamp.  Such a system would be characterized by an internal noise temperature that is lower than the ambient temperature, without the use of cryogenics.

The current experimental configuration also allows us to leverage the benefits of standard antennas when measuring ambient signal fields. Most prior work with Rydberg sensors has focused on using the atoms in free space. While this has several advantages (e.g. absolute accuracy, THz frequency detection, and low field perturbation), these sensors suffer from poor coupling to any particular RF mode, and the spectral response cannot be easily tuned or narrowed. By instead in-coupling with an external antenna we are able to select a well-defined RF mode, utilizing the antenna's spectral selectivity and gain. In this paradigm, the Rydberg sensor replaces the back-end of a receiver system, providing readout that can beat thermal Johnson noise limits and ease some impedance matching issues of traditional electronic sensors.

In parallel to increasing the RF-atom coupling, another critical area of research is to study alternate probing schemes for the thermal Rydberg atoms.  It is now well established that the common, continuous electromagnetically induced transparency (EIT) method has numerous limitations relative to the standard quantum limit (SQL) for an ideal quantum sensor \cite{meyer_assessment_2020}.  With better probing schemes, several orders of magnitude in sensitivity and bandwidth may be gained.  Additionally, adding a build-up cavity to recycle power in the currently expensive \SI{480}{\nano\meter} laser may be a route to improved performance and/or lower cost and size.

It is an exciting prospect for Rydberg RF sensors to become a useful piece of technology in the near future.  We must highlight that other physical platforms, such as electro-optics \cite{ savchenkov_photonic_2014,soltani_efficient_2017},  acousto-optics \cite{shao_microwave--optical_2019}, opto-mechanics \cite{forsch_microwave--optics_2020, jiang_efficient_2020}, and other photonic platforms are making corresponding advances.  In the longer term, Rydberg quantum sensors may be optimally suited to provide a full merger of classical and quantum communications.  Current experiments to achieve quantum frequency conversion \cite{han_coherent_2018, suleymanzade_tunable_2020} and coupling of Rydberg atoms to superconducting resonators \cite{hogan_driving_2012, morgan_coupling_2020}, are also leading in this important direction. Much foundational study is still required to discern how these quantum tools will mature to solve real-world problems.

\begin{acknowledgments}
We thank Fredrik Fatemi, Donald Fahey, and Jonathan Hoffman for helpful discussions.  This  work  was  partially  supported  by  the Defense Advanced Research Projects Agency (DARPA).
\end{acknowledgments}

\appendix
\section{Experimental details}
\label{app:ExpDetails}
For data presented, the atom chamber was heated to approximately \SI{50}{\degreeCelsius}, with higher temperatures leading to large optical depth, and decrease sensitivity. The measured optical depth of the ensemble at the $F=3\rightarrow F'=4$ $^{85}$Rb D2 transition was 2.4. The optical homodyne readout in our experiment is the same as that described in Ref.~\onlinecite{meyer_assessment_2020}.  Overall path length fluctuations in the homodyne are detected and stabilized using a colinear off-resonant beam, that is measured in heterodyne simultaneously with the balanced photodetector (Fig.~\ref{fig:main}).  The path is actively stabilized using an electro-optic modulator and a phase lock.  The optical powers are actively stabilized using acouto-optic modulators.

The \SI{480}{\nano\meter} laser is Pound-Drever-Hall locked to a stable reference cavity. The Rydberg coupling beam has a $1/e^2$ beam diameter of \SI{380}{\micro\meter} and a typical power of \SI{\sim500}{\milli\watt} at the atoms' location. The \SI{780}{\nano\meter} probe laser has a $1/e^2$ beam diameter of \SI{410}{\micro\meter} and is offset phase locked to a separate ``master'' laser referenced to rubidium spectroscopy. 
For most of the presented data, the total power in the optical homodyne/heterodyne probing beam was \SI{67}{\micro\watt} with \SI{17}{\percent} of the power in the homodyne probe sideband (\SI{11.6}{\micro\watt}). The optical LO power (not to be confused with the RF LO) was \SI{2.0}{\milli\watt}. The data of Figures \ref{fig:SensPerform} and \ref{fig:Signals} used a total power of \SI{22.3}{\micro\watt} (\SI{3.9}{\micro\watt} in the probing sideband) and an optical LO power of \SI{1}{\milli\watt}. For the microwave homodyne measurements in Figure \ref{fig:MWreadout}, the optical probe sidebands were turned off and the carrier frequency was moved to the probing resonance. The probing power was \SI{4.5}{\micro\watt}.

\section{Resonant and Off-Resonant Rydberg Response}
\label{app:RydResponse}

The response of the Rydberg sensor to arbitrary RF frequencies is described in detail in Reference \onlinecite{meyer_assessment_2020}. Here we provide a brief summary.

Far from resonance, the Stark shift of the target Rydberg state (i.e. the optically probed Rydberg state) depends on the atomic polarizability:
\begin{equation}\label{eq:offRes}
    \Omega_\text{off-res} = -\frac{1}{2}\alpha \left<E^2\right>_\tau
\end{equation}
and is proportional to the rms of the square of the total field amplitude. In this Article, we optically address the $\ket{59D_{5/2}}$ Rydberg state in rubidium 85.  For quasi-DC fields, the polarizability of this state is $\alpha=\SI{727.7}{\mega\hertz\centi\meter\squared\per\volt\squared}$ \cite{*[{This and all other Rydberg state properties were calculated using the ARC software package: }] [{}] sibalic_arc:_2017}.

Near a resonance, the Stark shift takes the form of an Autler-Townes splitting:
\begin{equation}\label{eq:onRes}
    \hbar\Omega_\text{on-res} = \wp \left<E\right>_\tau
\end{equation}
where $\wp$ is the transition dipole matrix element and $\left<E\right>$ is the rms amplitude of the RF field.  For the $\ket{59D_{5/2}}$ to $\ket{60P_{3/2}}$ transition at \SI{10.223336}{\giga\hertz}, $\wp=2210.6\,ea_0$. For the $\ket{59D_{5/2}}$ to $\ket{58F_{7/2}}$ transition at \SI{11.225754}{\giga\hertz}, $\wp=2211.3\,ea_0$.

The addition of an RF local oscillator field is an effective method for improving sensitivity of Rydberg sensors \cite{gordon_weak_2019,jing_atomic_2020,jau_vapor-cell-based_2020}, especially in the off-resonant square-law regime. Taking the Signal field as $E_\text{S}\cos((\omega+\delta)t)$ and the LO field as $E_\text{LO}\cos(\omega t - \varphi_\text{LO})$, we can derive the atomic response to the Signal field, given the presence of the LO.

Far from resonance, the Rydberg sensor acts as a square-law sensor. The squared total field becomes
\begin{align}
    \begin{split}
        E_\text{tot}^2 ={}& E_\text{S}^2\cos^2((\omega+\delta)t) + E_\text{LO}^2\cos^2(\omega t - \varphi_\text{LO}) \\
        & + 2E_\text{S}E_\text{LO}\cos((\omega+\delta)t)\cos(\omega t - \varphi_\text{LO})\nonumber
    \end{split}\\
    \begin{split}\label{eq:E2}
        ={}& E_\text{S}^2\cos^2((\omega+\delta)t) + E_\text{LO}^2\cos^2(\omega t - \varphi_\text{LO}) \\
        & + E_\text{S}E_\text{LO}(\cos(\delta t + \varphi_\text{LO})+\cos((2\omega+\delta)t - \varphi_\text{LO}))
    \end{split}
\end{align}
Taking a time average with the assumption that $\omega\gg\delta$ with $\delta$ less than the instantaneous bandwidth ($2 \pi f_\text{BW} = 1/\tau$), we get
\begin{equation}
    \left<E_\text{tot}^2\right>_\tau \approx E_\text{S}^2/2 + E_\text{LO}^2/2 + E_\text{S}E_\text{LO}\cos(\delta t + \varphi_\text{LO})
\end{equation}
Combined with Eq. \ref{eq:offRes}, we find a beat at frequency $\delta$ in the Stark shift that can be measured spectroscopically.
Note that the beat signal is linear in both the Signal and LO fields, meaning the Rydberg response is linear in the Signal with heterodyne gain from the LO.

Near resonance, in the Autler-Townes regime, the Rydberg sensor is linear in the field amplitude. We can find $\left<E_\text{tot}\right>$ by taking the root mean square of $E_\text{tot}$ and assuming that $E_\text{S}\ll E_\text{LO}$ to obtain
\begin{multline}
    \left<E_\text{tot}\right>_\tau=\sqrt{\left<E_\text{tot}^2\right>_\tau}
    \approx E_\text{LO} \langle \cos^2(\omega t - \varphi_\text{LO}) \\
    + \frac{E_\text{S}}{E_\text{LO}}(\cos(\delta t + \varphi_\text{LO})+\cos((2\omega+\delta)t - \varphi_\text{LO}))\rangle_\tau^{1/2}
\end{multline}
Again taking the time average with the assumption that $\omega\gg\delta$ and $\delta$ less than the instantaneous BW, we get
\begin{equation}
    \langle E_\text{tot}\rangle_\tau\approx \frac{E_\text{LO}}{\sqrt{2}}+\frac{E_\text{S}}{\sqrt{2}}\cos(\delta t + \varphi_\text{LO})
\end{equation}
Like the off-resonant case, there exists a beat at frequency $\delta$ in the Stark shift, with amplitude linear in $E_\text{S}$, that can be measured spectroscopically. Unlike the off resonant signal, the LO field does not amplify the beat signal. The benefit to adding an LO lies in biasing the Autler-Townes splitting away from zero where the small Stark shifts cannot be resolved within the linewidth of the spectroscopic signal.

\section{Readout sensitivity}
\label{app:ReadoutSens}

\begin{figure}[t]
\centering
\includegraphics[width=\columnwidth]{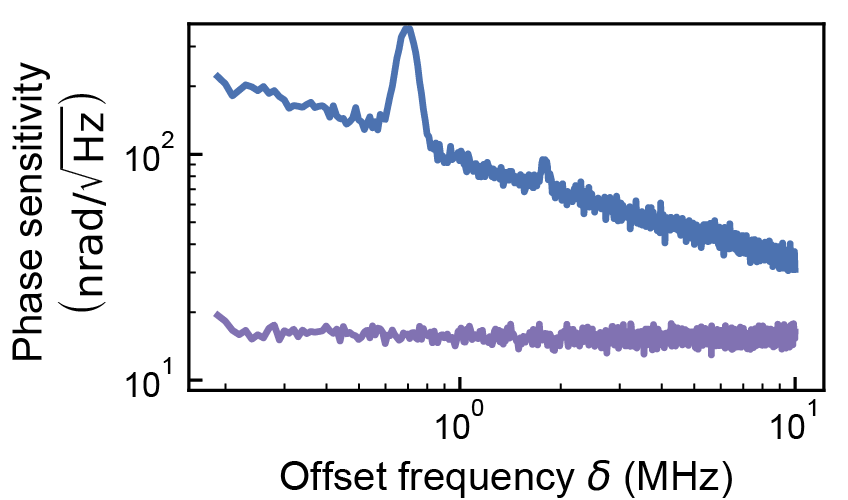}
\caption{Homodyne readout sensitivity.  The measured readout phase resolution, due to photon shot noise is shown in purple.  Additional laser phase noise leads to poorer overall resolution, shown in blue.}
\label{fig:readoutSens}
\end{figure}

Figure \ref{fig:readoutSens} presents a measurement of the optical homodyne phase sensitivity.  The photon shot noise level is \SI{2}{\nano\radian\per\sqrthz}, shown in purple. The total readout sensitivity is shown in blue, including an additional 1/$\sqrt{f}$ component due to laser phase noise. 

These measurements were performed with no RF or Rydberg atoms and are consistent with the sensor noise when atoms and signals are present.  

While the relative optical path length difference in our system is actively stabilized, our sensitivity to the probe laser phase noise is the result of unbalanced absolute path lengths between the optical probe and optical LO. Further engineering efforts to reduce this noise in the optical homodyne are readily possible by either reducing the overall size of the experiment (thereby improving the path length imbalance) and/or using a probing laser with narrower linewidth (thereby reducing the source laser phase noise).

\section{Waveguide performance}
\label{app:WaveguidePerf}

The microwave waveguide circuit is constructed from Rogers 3003 dielectric, 0.060" thick with \SI{35}{\micro\meter} thick copper plating on both sides. The waveguide slot is expanded at the atom location with the intent to increase the evanescent mode area to approximately match the size of the laser-induced ensemble.  Further, the waveguide gaps are asymmetric, to encourage localization of the electric field lines on one side of the trace. Independent DC bias voltages are applied to the central signal trace of the waveguide and the ground plane opposite the dielectric from the waveguide. The signal trace DC bias is applied using a Bias-Tee external to the vacuum chamber on the Signal input. These bias fields are used to cancel ambient electric fields, with the backplane biased to \SI{14.5}{\volt} and the signal trace at \SI{2.2}{\volt}, for all data shown.

\begin{figure}[t]
\centering
\includegraphics[width=\columnwidth]{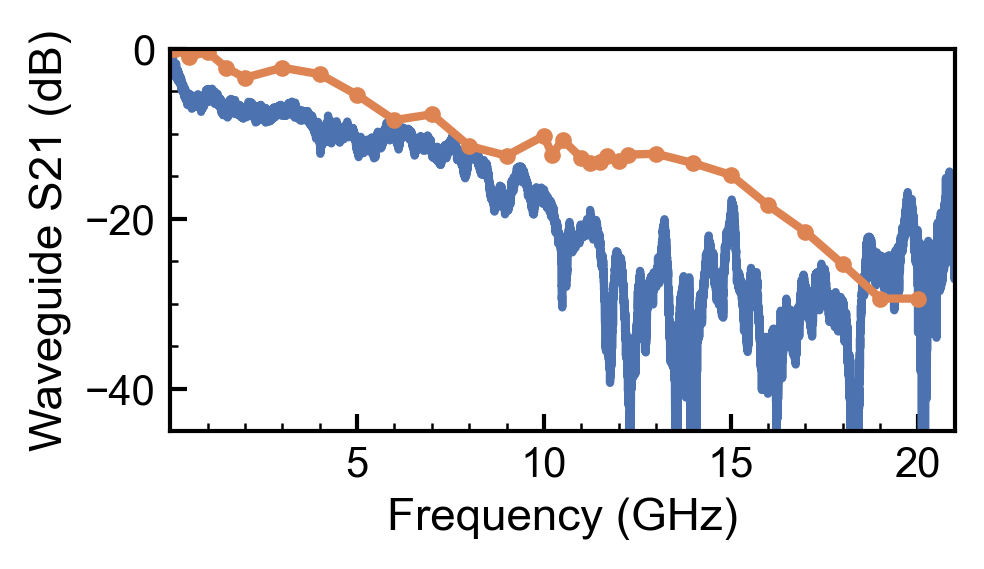}
\caption{In-situ measured (blue) and COMSOL predicted (orange) S21 of the coplanar waveguide. }
\label{fig:waveguideS21}
\end{figure}

We performed finite element multiphysics modelling of the waveguide using COMSOL Multiphysics modelling software. Analysis includes s-parameters and RF fields within and above the waveguide. The modelled s-parameter, S21, for the waveguide is shown in orange on Fig.~\ref{fig:waveguideS21}.  Experimentally, we observed a reduction in the measured board S21 over the 7-month operational period (Feb. 2020 to Sept. 2020), presumably due to infection of rubidium into the substrate (visually evidenced by discoloration of the substrate).  Initially, the measured S21 matched or slightly outperformed, the COMSOL prediction from 0 to 10~GHz.  The measured S21 at the end of experiments is shown in Fig.~\ref{fig:waveguideS21} in blue.  All reported Sensitivity measurements are referred to the waveguide input; we do not subtract waveguide losses to calculate the intrinsic Sensitivity.

We also use the COMSOL model to provide an empirical model of the waveguide conversion between input power and evanescent electric field amplitude over the waveguide gap in the region where the Rydberg atoms are excited. This conversion takes the form of:
\begin{equation}\label{eq:convFactor}
    E_\text{RF}=\sqrt{2PR_L} 10^{(P_\text{in}+\alpha\cdot f_\text{RF}+\beta+\zeta)/20}
\end{equation}
where $\sqrt{2PR_L}=\SI{0.316}{\volt\per\meter}$, $P_\text{in}$ is the input RF power at the input connector (in \si{\dBm}), $\alpha\cdot f_\text{RF}+\beta$ is the empirical model fit from COMSOL modelling of the waveguide, and $\zeta=\SI{-14}{\deci\bel}$ is a free parameter that accounts for degradation of the board performance relative to the COMSOL model. The fit parameters are $\alpha=\SI{0.69(12)}{\deci\bel\per\giga\hertz}$ and $\beta=\SI{46.4(13)}{\deci\bel}$.
This conversion is used in the Floquet theory comparison, described in the next section.

\section{Floquet theory}
\label{app:Floquet}
Predictions of the response of the Rydberg atoms to an arbitrary frequency are calculated using the Floquet theory described in Ref.~\onlinecite{meyer_assessment_2020}. The output of this model produces the expected Stark shift of a spectroscopically probed Rydberg state due to the presence of an arbitrary RF field frequency and amplitude. We estimate the expected signal output by calculating the dynamic Stark shift, $\Omega_\text{S}$, due to the beating of $E_\text{LO}$ and $E_\text{S}$.
This is converted to a beat signal by
\begin{equation}
    \text{Output}=20\cdot\log_{10}\left(\frac{\Omega_\text{S}}{D  \eta\sqrt{2P R_L}}\right)
\end{equation}
where $P=\SI{1}{\milli\watt}$, $R_L=\SI{50}{\ohm}$, $D=\lambda_p/\lambda_c$ is the Doppler scaling factor \cite{meyer_digital_2018}, and $\eta=\SI{2}{\mega\hertz\per\milli\volt}$ is a conversion factor between Stark shift and optical homodyne output voltage, as determined from the data shown in Figure \ref{fig:SensorSigs}(a)ii.

For the modelled Sensitivity, the Signal and LO powers are converted to electric field strength at the atoms using Eq.~\ref{eq:convFactor}. 
The resulting model of intrinsic Sensitivity (purple line in Fig.~\ref{fig:main}), corroborates the qualitative sensor response versus frequency.

\bibliography{CPW-Rydberg}

\end{document}